\documentstyle[prd,aps,eqsecnum,epsf]{revtex}
\begin{document}
\draft
\twocolumn[\hsize\textwidth\columnwidth\hsize\csname
@twocolumnfalse\endcsname
\title{PREHEATING AND REHEATING IN INFLATIONARY COSMOLOGY: A
PEDAGOGICAL SURVEY}
\vskip 1cm
\author{{\bf D. Boyanovsky$^{(a)}$, 
H.J. de Vega$^{(b)}$, R. Holman$^{(c)}$ and J. F. J. Salgado$^{(b)}$ }}   
\address
{ (a)  Department of Physics and Astronomy, University of
Pittsburgh, Pittsburgh, PA. 15260, U.S.A. \\
 (b)  LPTHE, \footnote{Laboratoire Associ\'{e} au CNRS UA280.}
Universit\'e Pierre et Marie Curie (Paris VI) 
et Denis Diderot  (Paris VII), Tour 16, 1er. \'etage, 4, Place Jussieu
75252 Paris, Cedex 05, France \\
 (c) Department of Physics, Carnegie Mellon University, Pittsburgh,
PA. 15213, U. S. A. }
\address{}
\date{August 1996}
\maketitle
\begin{abstract}
Recent progress in the preheating phenomena for inflationary
cosmology is reviewed. We first discuss  estimates of the preheating
time scale and  
particle production at the early stages of parametric amplification within the
Mathieu and Lam\'e approximations and we  analyze their precision and
limitations. The necessity of self-consistent calculations including
the non-linearity of the field theory  equations in an energy
conserving scheme is stressed. 
 The large $ N $  calculations including the field back-reaction are reviewed.
For spontaneously broken theories the issue of symmetry restoration is
analyzed. A discussion of the possibility and criterion for symmetry
restoration is presented. 
(To appear in the Proceedings of the Paris Euronetwork Meeting `String
Gravity', Observatoire de Paris, June 1996).
\end{abstract}
\vskip2pc]

\section{Introduction}

Research activity on inflationary cosmologies has continued steadily
since the concept of  inflationary cosmology was first proposed in
1981 \cite{guth}.  

It was recognized   that in order to merge an inflationary scenario
with standard 
Big Bang cosmology a mechanism to reheat the universe was needed. Such
a mechanism must 
be present in any inflationary model to raise the temperature of the
Universe at the end of inflation,  
thus the problem of reheating acquired further importance deserving
more careful investigation. 
The original version of reheating  envisaged that during the last
stages of  inflation when the  
accelerated universe
expansion slows down,  the energy stored in the oscillations of the
inflaton zero mode transforms into particles via single particle
decay. Such particle production 
reheats the universe whose temperature reduced enormously due to the
inflationary expansion \cite{newI}. 

It was realized recently\cite{frw,stb,kls,jap}, that in fact,
the elementary theory of reheating \cite{newI} does not describe accurately
the quantum dynamics of the fields.

\bigskip

Our programme on non-equilibrium dynamics of quantum field theory, started in
 1992\cite{nos1}, is naturally poised to provide a framework to study these 
problems. The larger goal of the program is to study the dynamics of
 non-equilibrium 
processes from a fundamental field-theoretical description, by solving
 the dynamical 
equations of motion of the underlying 
four dimensional quantum field theory for physically relevant problems: 
phase transitions out of equilibrium, particle production out of equilibrium,
 symmetry breaking and dissipative processes. 

The focus of our work is to describe the quantum field dynamics when
the  energy density is {\bf high}. That is, a large number of particles per
volume $ m^{-3} $, where $ m $ is the typical mass scale in the
theory. Usual S-matrix calculations apply in the opposite limit of low
energy density and since they only provide information on {\em in}
$\rightarrow$ 
{\em out} matrix elements,  are unsuitable for calculations of
expectation values. 
  Our methods were  naturally applied to different physical
problems like pion condensates \cite{dcc,nos3}, supercooled phase
transitions  \cite{nos1,nos2} and inflationary cosmology
\cite{frw,nos2,big,fut,fut2}.

An analogous  program has been pursued by the Los Alamos  group, whose
  research focuses on 
  non-linear effects in scalar QED linked to
the pair production in strong electric fields\cite{laCF}, 
 the Schwinger-Keldysh non-equilibrium formalism in the large $ N $
  expansion\cite{laNG}, 
and the dynamics of   chiral condensates in such framework\cite{dccla}.

\section{Preheating in inflationary universes}

As usual in inflationary cosmology, matter is described in an
effective way by a self-coupled scalar field $ \Phi(x) $ called the inflaton. 
The spacetime geometry is a cosmological spacetime with metric $ ds^2
= (dt)^2 - a(t)^2 \; (d{\vec x})^2 $, where $ a(t) $ is the scale factor.

The evolution equations for the $k$-modes of the inflaton field as 
considered by different groups can be summarized as follows,
\begin{equation}\label{modos}
{\ddot \chi}_k  + 3 H(t) \; {\dot \chi}_k + \left( {{k^2} \over {a^2(t)}} +
M^2[\Phi(.)] \right)  \chi_k(t) = 0
\end{equation}
where $ H(t) = {\dot a}(t)/a(t) , \; $ and $ M^2[\Phi(.)] $ 
is the effective mass felt by the modes. The
expression considered depends on the model (here the $\lambda \;
\Phi^4 $ model) and the approximations made.  The value of $\lambda$
is scenario-dependent but it is usually very small.
$ M^2[\Phi(.)] $
depends on the scale factor and  on the physical state. Therefore, it
depends on the modes $  \chi_k(t) $ themselves in a complicated
way. One is definitely faced with a complicated non-linear problem. We
call `back-reaction' the effect of the modes  $  \chi_k(t) $ back on
themselves through  $ M^2[\Phi(.)] $. 

In the initial stage, all  the energy is assumed in the zero mode of
the field $  \phi(t) $
\cite{nos1,nos2,nos3,big,stb,kls,kof,jap}. That is, the field expectation value
$ \phi(t) \equiv <\Phi(x)> $, where
$ <\cdots > $ stays for  the expectation value in the translational
invariant but non-equilibrium quantum  
state. For very weakly coupled theories, and early times, such that the
back-reaction effects of the non-equilibrium quantum fluctuations can
be neglected, one can  approximate,
\begin{equation} \label{prehM}
M^2[\Phi(.)] \simeq m^2 + { {\lambda} \over 2} \phi(t)^2
\end{equation}
At this moment, the scale factor is set to be a 
 constant in ref.\cite{stb,jap}. Refs.\cite{nos1,nos2,nos3,big} consider
 Minkowski spacetime from the start. In ref.\cite{kls},  the scale
 factor is set to be a   constant for a model without 
$ \lambda \Phi^4 $ inflaton self-coupling. That is, for the classical
potential\cite{kls}
\begin{equation}\label{Vines}
V = \frac12 m^2 \Phi^2 +  g  \sigma^2 \Phi^2 \; ,
\end{equation}
one can consider as classical solution $  \Phi(t) =  \Phi_0 \;
\cos(mt) , \; \sigma = 0 $. In such case, the Mathieu equation
approximation is exact in Minkowski spacetime.
However, the potential (\ref{Vines}) is
unstable under renormalization (a $ \Phi^4 $ counterterm is needed from
the one-loop level). Hence, the $ \lambda = 0 $ choice is a fine-tuning not
protected by any symmetry.
In a second case a massless selfcoupled $ \lambda \Phi^4 $ field in a
radiation dominated universe $ a(t) \propto \sqrt{t} $ is considered
in conformal time \cite{kls}. In such specific case the classical
field equations take the Minkowski form for $  \sqrt{t}  \Phi$.

In a way or another, using
the classical oscillating behaviour of $ \phi(t) $,  one is 
led by eq.(\ref{modos}) to an effective mass that oscillates in time. In
this approximation (which  indeed, may be very good for small
coupling, see \cite{big}), 
eq.(\ref{modos}) exhibits {\bf parametric resonance} as noticed first
in ref\cite{stb}. Namely, there are allowed and forbidden bands in $ k^2 $. The
modes within   the  forbidden bands grow exponentially whereas those 
in the allowed bands stay with bounded modulus. The growth of the
modes in the  forbidden bands translates into profuse particle production, the particles being created
with these particular unstable momenta. The rate of particle production is determined by the
imaginary part of the Floquet index in these unstable bands. 
Notice that the approximation (\ref{prehM}) breaks down as
soon as many particles are produced. Namely, when the energy of the
produced particles becomes of the order of the zero-mode energy and
eq.(\ref{prehM}) is no more valid. 

Now, in order to compute quantitatively the particles produced one
needs the form of  $ \phi(t) $. In ref.\cite{stb,kls,jap} $ \phi(t) $
is approximated by a cosinus function in the calculations. 
The mode equations become a Mathieu equation (the scale factor is set
to be a constant).  In ref.\cite{stb} the Bogoliubov-Krylov approximation
is used to compute estimates. In  ref.\cite{kls,kof}, estimates are
obtained using asymptotic formulas for the Mathieu equation. In
ref.\cite{big} the exact classical solution is used (a cn Jacobi
function) to compute estimates. 

Let us now compare the results from the exact mode solutions obtained
in ref.\cite{big} with the Mathieu equation approximation to it.
In units where $ m^2 = 1 $ and setting $ \eta(t) \equiv \sqrt{
{\lambda} \over 2} \; \phi(t) $, one finds
\begin{eqnarray}\label{etac}
\eta(t) &=& \eta_0\; \mbox{cn}\left(t\sqrt{1+\eta_0^2}, {\bar k}\right)
\cr \cr
 {\bar k} &=& {{\eta_0}\over{\sqrt{2( 1 +  \eta_0^2)}}}\; , 
\end{eqnarray}
where cn stands for the Jacobi cosinus and we choose for initial
conditions $ \eta(0) =  \eta_0\; , \; {\dot \eta}(0) = 0 $.

Here $ \eta(t) $ has period $
4 \omega \equiv {{ 4 \, K( {\bar k})}\slash {\sqrt{1+\eta_0^2}}} $,
where $ K( {\bar k}) $ is
the complete elliptic integral of first kind, and $  \eta(t)^2 $
has period $ 2  \omega$.

  Inserting this form for
$\eta(\tau)$ in eqs.(\ref{prehM}) and (\ref{modos}) yields
\begin{equation}\label{modsn}
 \left[\;\frac{d^2}{dt^2}+k^2+1+   \eta_0^2\;
\mbox{cn}^2\left(t\sqrt{1+\eta_0^2}, {\bar k}\right) \;\right]
 \chi_k(t) =0 \; . \label{nobackreaction}
\end{equation}
This is the Lam\'e equation for a particular value of the coefficients that
make it solvable in terms of Jacobi functions \cite{herm}. 
As shown in ref.\cite{big}, this equation has only one forbidden band
for positive $ k^2 $ going from   $ k^2 = 0 $ to  $ k^2 =
{{\eta_0^2}\over 2} $. One can choose Floquet  solutions of
eq.(\ref{nobackreaction}) that fullfil  the relation
\begin{equation}\label{floq}
 U_k(t + 2  \omega) =   e^{i F(k)} \; U_k(t),
\end{equation}
where the Floquet indices $ F(k) $ are independent of $t$. In the
forbidden band the  $ F(k) $ posses an imaginary part. 
The production rate is determined by the imaginary 
part of the Floquet index.

The exact form of  $ F(k) $ results \cite{big},
$$
 F(k) = -2 i K( {\bar k}) \;  Z(2  K( {\bar k}) \,v)  +  \pi
$$
 where $ Z(u) $ is the Jacobi zeta function \cite{erd} and
$ v $ is a function of $ k $ in the forbidden band defined by 
\begin{equation}\label{qprohi}
k =  {{\eta_0}\over {\sqrt2}}\, \mbox{cn}(2  K( {\bar k})\,v,k) \; ,
\; 0 \leq v \leq \frac{1}{2}. 
\end{equation}
All these elliptic functions posses fastly convergent expansions in
powers of the elliptic nome 
$$
 q \equiv  e^{-\pi K'( {\bar k})/ K( {\bar k})} \; .
$$ 
  Since $ 0 \leq {\bar k}  \leq 1/\sqrt2 $ [see eq.(\ref{etac})], we have
\begin{equation}\label{qsomb}
 0 \leq  q  \leq e^{-\pi} = 0.0432139\ldots \; . 
\end{equation}
Then, 
\begin{equation}
F(k) = 4i\, \pi  \;  q \; \sin(2\pi v)\;\left[ 1 + 2 \, q
\; \cos2\pi v   + O( q^2)\right]  + \pi \; .\label{easyfloquet}
\end{equation}
The imaginary part of this function has a maximum at $ k = k_1 =
 \frac12 \; \eta_0 \; (1 -  q )  + O( q^2)  $ where \cite{big}
\begin{equation}\label{Flame}
{\cal F} \equiv  Im F(k_1) = 4\, \pi \; q + O(q^3) \; .
\end{equation}
This simple formula gives the maximum  of the imaginary part of the
Floquet index  in the forbidden band with a
precision better than $ 8.\, 10^{-5} $. $ q $ can be expressed in
terms of $ \eta_0 $ as follows \cite{big}
$$
q  =  \frac12 \;  {{ (1+\eta_0^2)^{1/4} -  (1+\eta_0^2/2)^{1/4}}
\over { (1+\eta_0^2)^{1/4} +  (1+\eta_0^2/2)^{1/4}}}  \; .
$$
with an error smaller than $\sim 10^{-7} $.

Let us now proceed to the Mathieu equation analysis of this
problem. The cn Jacobi function can be expanded as \cite{gr}
$$
{\rm cn} (z, {\bar k}) = (1-q) \cos(1-4q)z + q \cos3z + O(q^2) \; .
$$
To zeroth order in $ q $ we have
$$
\eta(t)^2 = {{\eta_0^2}\over 2} \left[ 1 + \cos(2t\sqrt{1 + \eta_0^2})
\right] + O(q) \; .
$$
and $ 2 \omega = \pi/\sqrt{1 + \eta_0^2} + O(q) $.
Under such approximations eq.(\ref{modsn}) becomes the Mathieu
equation \cite{abr}
\begin{equation}\label{mathieu}
{{d^2 y}\over {dz^2}} + \left( a - 2  {\bar q} \cos2z \right)y(z) = 0
\; ,
\end{equation}
where
$$
a = 1 + {{k^2 -  {{\eta_0^2}\over 2} }\over {\eta_0^2 + 1 }} \; , \; 
 {\bar q} = {{\eta_0^2}\over{ 4(\eta_0^2  + 1)}}  
$$
and $ z = \sqrt{ \eta_0^2 + 1 } \; t $. Notice that $ 0 \leq   {\bar q}
\leq 1/4 $ in the present case. Eq.(\ref{mathieu}) posses an infinite
number of forbidden bands for $ k^2 > 0 $. The  lower and upper edge of the first band are
respectively\cite{abr}
$$
k^2_{inf} =   {{\eta_0^2}\over 4}\left[ 1 -  {{\eta_0^2}\over{
2^5(\eta_0^2 + 1)}}  +  {{\eta_0^4}\over{ 2^{10}(\eta_0^2 + 1)}} + \ldots
\right] \; ,
$$
till
$$
k^2_{sup} =   {{\eta_0^2}\over 4}\left[ 3 -  {{\eta_0^2}\over{
2^5(\eta_0^2 + 1)}}  -  {{\eta_0^4}\over{ 2^{10}(\eta_0^2 + 1)}} + \ldots
\right] \; .
$$

These values must be compared with the exact result for the Lam\'e
equation (\ref{nobackreaction}) : $ k^2_{inf} = 0 \; , \; k^2_{sup} =
{{\eta_0^2}\over 2} $. The width of the band is well approximated but
not its absolute position. The numerical values of the maximum of the
imaginary part of the Floquet index are given in Table I
and compared with the exact values from eq.(\ref{Flame}). 

We see that the Mathieu approximation {\bf underestimates} the exact result
by a fraction ranging from $13$\% to $ 39$\%. The second forbidden
band in the Mathieu equation yields $  {\cal F}_{ Mathieu} =
0.086\ldots $ for  $ \eta_0 \to \infty $. This must be compared with $
{\cal F}_{Lame} = 0 $ (no further forbidden bands). 

In ref.\cite{kof}, an even  larger
discrepancy between Lame and Mathieu
Floquet indices has been reported within a different approximation scheme.

It must be noticed that the Floquet indices entering in exponents 
produces a dangerous error propagation. For example, the number of
particles produced during reheating is of the order  of the exponential
of $ 2  {\cal F} $ times the reheating time in units of $ \pi/\sqrt{ 1 +
{\eta_0^2}} $. An error of $25$\% in $   {\cal F} $ means an error of
$25$\% in the exponent. (For instance, one would find $10^9$ instead
of $10^{12}$). 
 
The mode equations (\ref{modos}) apply to the self-coupled $ \lambda \;
\Phi^4 $ scalar field. Models for reheating usually contain at least
two fields: the inflaton and a lighter field $ \sigma(x) $ in which
the inflaton decays. For a $ g  \sigma^2 \Phi^2 $ coupling, the  mode
equations for the  $ \sigma $ field take the form \cite{nos3,kls,jap}
\begin{equation}\label{modsi}
{\ddot V}_k  + 3 H(t) {\dot V}_k + \left( {{k^2} \over {a^2(t)}} +
m_{\sigma}^2 + {g\over {\lambda}} F[\Phi(.),\sigma(.)] \right)  V_k(t) = 0
\end{equation}
A new dimensionless parameter $ { g \over  {\lambda}} $ appears
here. Neglecting the $ \sigma $ and $ \Phi $ backreaction, we have
\begin{equation}\label{preH}
 F[\Phi(.),\sigma(.)] \simeq   \eta^2(t) \; .
\end{equation}

In ref.\cite{nos2,big}, it is shown that abundant particle
production (appropriate for reheating) shows up  even for $ g =
\lambda$. 

\bigskip

\begin{table} \centering
\begin{tabular}{|l|l|l|l|}\hline
$ \eta_0 $ &  $ {\cal F}_{Lame} $  &  $ {\cal F}_{ Mathieu} $ &  
 $ \% error $ \\ \hline 
$ $&  $ $ & $ $ & $ $ \\
1 & $ 0.2258 \ldots $ & $ 0.20 \ldots $ & $ 13$\% \\
$ $&  $ $ & $ $ & $ $ \\ \hline
$ $&  $ $ & $ $  & $ $ \\
$ 4 $ & $ 0.4985\ldots $ & $ 0.37\ldots $ & $ 35$\% \\
$ $&  $ $ & $ $ & $ $ \\ \hline
$ $&  $ $ & $ $  & $ $ \\
 $ \eta_0 \to \infty $ &  $ 4\pi e^{-\pi} = 0.5430\ldots $ & $ 0.39\ldots
$ & $ 39$\% \\ $ $&  $ $ & $ $ & $ $ \\ \hline
\end{tabular}

\bigskip

\label{table1}
\caption{ The maximum of the imaginary part of the Floquet index ${\cal
F}$ for the Lam\'e  equation and for its Mathieu approximation.}
\end{table}

Eqs.(\ref{modsi})-(\ref{preH}) becomes a Lam\'e equation when $
\eta(t) $ is approximated by the classical solution in Minkowski
spacetime given by (\ref{etac}). Such Lam\'e equation is solvable in closed
form when the couplings $ g  $  and $ \lambda $  are related as
follows \cite{big}
$$
 {{2 g}\over {\lambda}} = n(n+1) \; , \; n=1,2,3,\ldots
$$
In those cases there are $ n $ forbidden bands for $ k^2 \geq 0
$. The  Lam\'e equation exhibits an infinite number of
forbidden bands for generic values of $ {g\over {\lambda}} $. 
The Mathieu and WKB approximations has been applied
in the non-exactly solvable cases \cite{kls,kof,jap}. However, as the
above analysis shows (see Table I)  such estimations 
 cannot be trusted quantitatively. The only available precise method
consists on  accurate numerical calculations as those of
ref.\cite{nos2,nos3,big} (where the precision is at least $ 10^{-6} $).
  
Estimates in the cosinus approximation for FRW-de Sitter backgrounds
and open universes
using the Bogoliubov-Krylov approximation are given in
ref.\cite{kaiser}. In ref.\cite{son}  the Bogoliubov-Krylov
approximation is applied to the large $ N $ equations.

Applications of preheating to various relevant aspects of the early
cosmology are considered in ref.\cite{apli}.

\bigskip

As soon as the quantum fluctuations grow and cease to be negligible
compared with the the classical piece (\ref{preH}), all the approximations
discussed so far (Lam\'e, Mathieu, etc.) break down. This time is the
so-called preheating time $ t_{reh} $ \cite{big}.
One can estimate  $ t_{reh} $ by equating the zero mode energy
(\ref{preH}) with the estimation of the quantum fluctuations derived
from the unstable Floquet modes \cite{kls,big}. Such estimation yields
when the Lam\'e Floquet indices are used
\cite{big},

\begin{equation}
t_{reh} \approx {1 \over B} \, \log{{N(1+\eta^2_0/2) \over { g \sqrt
B}}}\; , \label{maxtime} 
\end{equation}
where
\begin{eqnarray}\label{ByN}
B &=&  \displaystyle{
8\, \sqrt{1+\eta_0^2}\; { q } \; (1 - 4 { q }) +  O(
{ q }^3) }\; , \cr \cr
N &=& {4 \over {\sqrt{ \pi}}} \; \sqrt{   q }\;
{{ ( 4 + 3 \, \eta_0^2) \, \sqrt{  4 + 5 \, \eta_0^2}}\over{
 \eta_0^3 \, (1+\eta_0^2)^{3/4}}} \left[ 1 + O ( q )\right]\; \; .
\end{eqnarray}

Where $ B $ is determined by the maximum of the imaginary part of the
Floquet index in the 
band. It is now clear that a few percent correction to the Floquet
indices will result in a large error in the estimate of the preheating
time scale.  

\section{The end of preheating and beyond: self-consistent large $N$
calculations} 

In order to compute physical magnitudes  beyond  $ t_{reh} $, one {\bf
must} solve self-consistently the field equations including the back reaction.
In  ref.\cite{nos2,big} this is done for the $ N \to \infty $
limit and in ref.\cite{nos3}  to one-loop order. 

In the large $ N $ limit the zero mode and $k$-modes renormalized
equations take the form,

\begin{eqnarray}
& & \ddot{\eta}+ \eta+
\eta^3+ g \;\eta(t)\, \Sigma(t)  = 0 \label{modo0} \; , \\
& & \left[\;\frac{d^2}{dt^2}+k^2+1+
\;\eta(t)^2 + g\;  \Sigma(t)\;\right]
 \varphi_k(t) =0 \; , \label{modok} 
\end{eqnarray}
where $g \Sigma(t)$ is given by
\begin{eqnarray}
g \Sigma(t) & = & g \int_0^{\infty} k^2
dk \left\{ 
\mid \varphi_k(t) \mid^2 - \frac{1}{\Omega_k}
 \right. \nonumber \\
&   & \left. 
+ \frac{\theta(k-\kappa)}{2k^3}\left[ 
 -\eta^2_0 + \eta^2(t) + g \; \Sigma(t) \right] \right\} \; ,\cr \cr
{\Omega_k}&=& \sqrt{k^2+1 +  \eta^2_0} \; .
\label{sigmafin}  
\end{eqnarray}
We choose  the initial state such that at $t=0$ the quantum
fluctuations are in the ground  state of the oscillators.
That is,
$$
 \varphi_k(0) = {1 \over {\sqrt{ \Omega_k}}} \quad , \quad 
{\dot \varphi}_k(0) = - i \; \sqrt{ \Omega_k} \; ,
$$
$$
\eta(0) = \eta_0  \quad , \quad {\dot\eta}(0) = 0\; .
$$
In the one-loop approximation the term $  g \; \Sigma(t) $ is absent
in eq.(\ref{modok}).

Eqs.(\ref{modok})-(\ref{sigmafin})
were generalized to cosmological spacetimes (including
the renormalization aspects) in ref.\cite{frw}. 

Eqs.(\ref{modok}) is an infinite set of coupled non-linear
differential equations in the $k$-modes $ \varphi_k(t) $ and in the
zero mode $ \eta(t) $. We have numerically solved eqs.(\ref{modok})
for a variety of couplings and initial conditions
\cite{nos2,nos3,big}. 
In figs. 1-3 we display $  \eta(t), \;   \Sigma(t)$ 
and the total number of created particles $ N(t) $ as a function of $
t $ for $ \eta_0 = 4 $ and $ g = 10^{-12} $ \cite{big}. One sees that
the quantum fluctuations are indeed very small till we approach the
reheating time $ t_{reh} $. As one can see from the figures, 
eq.(\ref{maxtime})  gives a very good approximation for  $ t_{reh} $ and
for the behaviour of the quantum fluctuations for $ t \leq  t_{reh} $.
For times earlier than  $ t_{reh} $, the Lam\'e estimate for the
envelope of $ \Sigma(t) $  and  $ N(t) $ take the form \cite{big},
\begin{eqnarray}\label{polenta}
 \Sigma_{est-env}(t) &=& { 1 \over { N \, \sqrt{t}}}\; e^{B\,t}\;
  , \cr\cr
 N_{est-env}(t) &=& {1 \over {8 \pi^2}} \; {{4 + \frac92 \eta_0^2}
\over { \sqrt{4 + 5 \eta_0^2}}} \;   \Sigma_{est-env}(t)\; .
\end{eqnarray}
where $ B $ and $ N $ are given by eq.(\ref{ByN}).

This analysis presents a very clear physical picture of the properties of the ``gas'' of
particles created during the preheating stage. This ``gas'' turns to be {\bf out} of thermodynamic equilibrium as
shown by the energy spectrum obtained, but it still  fulfills a
radiation-like equation of state $ p \simeq e/3 $ \cite{big}.

In approximations that do not include the backreaction effects there
is infinite particle production, 
since these do not maintain energy conservation\cite{stb,jap}. 
The full backreaction problem and the approximation 
used in our work exactly conserves the energy.
As a result, particle production shuts-off when 
the back-reaction becomes important ending  the preheating
stage.

In ref. \cite{dort}   results  of ref.\cite{nos2} are rederived with a
different renormalization scheme.

\section{Broken symmetry and its quantum restoration}

Up to now we discussed the unbroken symmetry case $ m^2 = 1 > 0 $. In the
spontaneously broken symmetry case  $ m^2 = -1 < 0 $, the classical
potential takes the form
\begin{equation}\label{potcla}
V(\eta) = \frac14 (\eta^2 - 1 )^2
\end{equation}
In refs.\cite{nos2,big} the field evolution is numerically solved
in the large $ N $ limit for  $ m^2 < 0 $. Analytic estimations are
given in ref.\cite{big} for early times. 

In ref.\cite{nos2} it is shown that for small couplings and small
$ \eta_0 $, the order
parameter damps very quickly to a constant value close to the
origin. That is, the order parameter ends close to the classical false
vacuum and far from the classical minimum $\eta = 1$. The fast
damping is explained,   in this spontaneously broken symmetry case
by the presence of  Goldstone bosons. The  massless Goldstone
bosons dissipate very efficiently the energy from the zero mode. We
plot in figs. 4 and 5,  $  \eta(t) $ and $  \Sigma(t) $ for $ \eta_0 =
10^{-5} $ and $ g = 10^{-12} $ \cite{big}. The profuse particle
production here is due to spinodal unstabilities.

In ref.\cite{kls2}, it is claimed that quantum fluctuations have
restored the symmetry in the situations considered in ref.\cite{nos2}.
However, the presence of massless Goldstone bosons and the non-zero
limiting value of the order parameter clearly show that the symmetry {\bf is}
always broken in the cases considered in  ref.\cite{nos2,big}. These
results rule out the claim in ref.\cite{kls2}. In particular, the asymptotic value of the
order parameter results from a very detailed dynamical evolution that conserves energy. 

In the situation of  `chaotic initial conditions' but with a broken
symmetry tree level potential, the 
issue of symmetry breaking is more subtle. In this case the zero mode
is initially displaced with a 
large amplitude and very high in the potential hill. The total energy
{\em density} is non-perturbatively 
large. Classically the zero mode will undergo oscillatory behavior
between the two classical turning 
points, of very large amplitude and the dynamics will probe both
broken symmetry states. Even 
at the classical level the symmetry is respected by the dynamics in
the sense that the time evolution 
of the zero mode samples equally both vacua. This is not the situation
that is envisaged in   
usual symmetry breaking scenarios.
For broken symmetry situations there are no finite energy field
configurations that can 
sample both vacua. In the case under consideration with the zero mode
of the scalar field with very 
large amplitude and with an energy density much larger than the top of
the potential hill, there 
is enough energy in the system to sample both vacua. (The energy is
proportional to the spatial 
 volume). Parametric amplification transfers energy from the zero mode
to the quantum fluctuations. 
Even when only a fraction of the energy of the zero mode is
transferred thus creating a  
non-perturbatively large number of particles, the energy in the
fluctuations is very large, and the 
equal time two-point correlation function is non-perturbatively large
and the field fluctuations are 
large enough to sample both vacua. The evolution of the zero mode is
damped because of this 
transfer of energy, but in most generic situations it does not reach
an asymptotic time-independent 
value, but oscillates around zero, sampling the tree level minima with
equal probability. 
This situation is reminiscent of finite temperature in which case the
energy density is finite and above 
a critical temperature the ensemble averages sample both tree level
vacua with equal probability 
thus restoring the symmetry. In the  dynamical case, the ``symmetry
restoration'' is just a consequence 
of the fact that there is a very large energy density in the initial
state, much larger than the top of the 
tree level potential, thus under the dynamical evolution the system
samples both vacua equally.   
This statement is simply the dynamical equivalent of the equilibrium
finite temperature statement that 
the energy in the quantum fluctuations is large enough that the
fluctuations can actually sample both 
vacua with equal probability. 

Thus the criterion for symmetry restoration  when the
tree level potential allows 
for broken symmetry states is that the energy density in the initial
state be larger than the top of the 
tree level potential. That is when the amplitude of the zero mode is
such that $ V(\eta_0) > V(0) $.  
In this case the dynamics will be very similar to the unbroken
symmetry case, the amplitude of the 
zero mode will damp out, transferring energy to the quantum
fluctuations via parametric amplification, 
but asymptotically oscillating around zero with a fairly large amplitude.

To illustrate this point clearly, we plot in figs. 6 and 7,   $
\eta(t) $ and $  \Sigma(t) $ for $ 
\eta_0 = 1.6 > \sqrt2 $ [and hence $  V(\eta_0) > V(0) $, see
eq.(\ref{potcla})] and $ g = 10^{-3} $. We find the typical behaviour
of unbroken symmetry. Notice again that the effective or tree level
potential is an irrelevant 
quantity for the dynamics, the asymptotic amplitude of oscillation of
the zero mode is $\eta \approx 
0.5$, which is smaller than the minimum of the tree level potential
$\eta=1$ but the oscillations are 
symmetric around $\eta=0$. 

Since the dynamical evolution sampled both vacua symmetrically  from the
beginning,  there never was a symmetry breaking in
the first place, and ``symmetry restoration''   is just the statement
 that the initial state has enough 
energy such that the {\em dynamics}  probes  both vacua symmetrically
 despite the fact that the 
tree level potential allows for broken symmetry ground states.

In their comment (hep-ph/9608341), KLS seem to agree with our
conclusion that in the situations studied in our articles\cite{nos2,nos3}, 
when the expectation value is
released below the potential hill, symmetry restoration does not occur.

We will present a deeper analytical and numerical study of the subtle
aspects of the dynamics in 
this case in a forthcoming article\cite{fut}.

\section{Linear vs. nonlinear dissipation (through particle creation)}

As already stressed, the field theory dynamics is unavoidable nonlinear for 
processes  like preheating and reheating. It is however interesting to
study such processes in the amplitude expansion. This is done in
detail in refs.\cite{nos2,nos3}. To dominant order, the amplitude
expansion means to linearize the zero mode evolution equations. This
approach permits an analytic resolution of the evolution in closed
form by Laplace transform. Explicit integral representations for $
\eta(t) $ follow as functions of the initial data
\cite{nos2,nos3}. Moreover, the results can be clearly described in
terms of S-matrix concepts (particle poles, production thresholds,
resonances, etc.). 

Let us consider  the simplest  model where the inflaton $\Phi$ couples
to a another scalar $\sigma$ and to a fermion field $\psi$, and
potential\cite{nos3} 
\begin{eqnarray}\label{mod2c}
V &=& {1 \over 2} \left[  m^2_{\Phi}\Phi^2 +  m_{\sigma}^2 \sigma^2 
+ g \; \sigma^2 \Phi^2  \right] \cr \cr
&+&{ \lambda_{\Phi} \over 4!}  \, \Phi^4 
     +{ \lambda_{\sigma} \over 4!}  \, \sigma^4+
{\bar{\psi}} (m_{\psi} + y \Phi  ) \psi \;. \nonumber
\end{eqnarray}
In the  unbroken symmetry  case $ (m^2_{\Phi} > 0) $  the inflaton is
always stable and we found for the order parameter (expectation value
of $ \Phi$) evolution  in the amplitude expansion \cite{nos3},
\begin{eqnarray}
\eta(t)&=&\frac{\eta_i}
{1-\frac{\partial\Sigma(i m_{\Phi})}{\partial  m_{\Phi}^2}}  
\cos [m_{\Phi} t]\cr \cr 
&+&{{2\, \eta_i }\over{\pi}} \int_{ m_{\Phi}+ 2m_{\sigma}}^{\infty}
{{\omega\Sigma_I(\omega) \cos\omega
t\;d\omega}\over{[\omega^2- m_{\Phi}^2-
\Sigma_R(\omega)]^2+ \Sigma_I(\omega)^2}}\;. \label{stable}
\end{eqnarray}
where $\Sigma_{\rm physical}(i\omega\pm 0^+)=\Sigma_R(\omega)\pm
i\Sigma_I(\omega)$ is the inflaton self-energy in the physical sheet,
$  {\eta_i} = \eta(0) $ and $ {\dot  \eta}(0) = 0 $. The first term is
the contribution of the one-particle pole (at the physical inflaton
mass). This terms oscillates forever with constant amplitude. The
second term is the cut contribution $ \eta(t)_{cut} $
corresponding to $ \Phi \to  \Phi + 2 \sigma $. 

In general, when
$$
\Sigma_I(\omega\to\omega_{\rm{threshold}}) \buildrel
{\omega\to\omega_{\rm{threshold}} } \over = B \;
(\omega-\omega_{\rm{threshold}})^\alpha \; ,
$$ 
the the cut contribution behaves  for late times as
\begin{eqnarray}
\eta(t)_{cut} &\simeq & {{2\, \eta_i }\over{\pi}}
{{ B  \; \omega_{\rm{threshold}} \; \Gamma(1+\alpha)}\over 
{[\omega^2- m_{\Phi}^2- \Sigma_R(\omega_{\rm{threshold}})]^2}}\cr \cr
&&
t^{-1-\alpha} \cos\left[\omega_{\rm{threshold}} t +
 \frac{\pi}2 (1+\alpha) \right] \; .
\end{eqnarray}
Here, $ \omega_{\rm{threshold}} =  m_{\Phi} +2  M_\sigma $ and 
$ \alpha = 2 $ since to two-loops,\cite{nos3} 
$$
\Sigma_I(\omega)\buildrel{\omega \to {m_{\Phi} + 2  M_{\sigma}}}\over= 
\frac{2 g^2 \pi^2}{(4\pi)^4} 
\frac{M_\sigma \sqrt{ m_{\Phi}}}{( m_{\Phi}+2 M_\sigma)^{7/2}}
[\omega^2-( m_{\Phi}+2 M_\sigma)^2]^2\;.
$$

In the broken symmetry  case $ (m^2_{\Phi} < 0) $  we may have either $
M < 2 m_{\sigma}$ or $ M > m_{\sigma} $, where $ M $ is the
physical inflaton mass. ( $ M = |  m_{\Phi}| \sqrt2 $ at the tree
level). In the first case the inflaton is stable and eq.(\ref{stable})
holds. However, the self-energy starts now at one-loop and vanishes at
threshold with  a power $ \alpha = 1/2 $. For  $ M > m_{\sigma} $
the inflaton decomes a resonance (an unstable particle) with width
(inverse  lifetime)
$$
\Gamma =  {g^2\Phi^2_0\over{8\pi M}}\sqrt{1-{{4
m^2_{\sigma}}\over{M^2}}}  \; .
$$
This pole dominates $ \eta(t) $ for non asymptotic times
\begin{equation}\label{BW}
\delta(t)\simeq\delta_i\, A\; e^{-{\Gamma t/2}}\;
\cos(Mt+\gamma) \; , 
\end{equation}
where
\begin{equation}
A=1+ {{\partial\Sigma_R(M)}\over{\partial M^2}}
\;,\quad\quad
\gamma= -{{\partial\Sigma_I(M)}\over{\partial M^2}}\;.\nonumber
\end{equation}
In summary, eq.(\ref{BW}) holds provided: a) the inflaton is a resonance
and b)  $ t \leq\Gamma^{-1}\ln(\Gamma / M_\sigma)$. 
For later times the fall off is with a power law $t^{-3/2}$
determined by the spectral density at threshold as before\cite{nos3}.

In ref.\cite{nos3} the selfconsistent nonlinear evolution is computed
to  one-loop level for the model (\ref{mod2c}). In fig. 8 $ \eta(t) $
is plotted as a function of time for $ \lambda = g = 1.6 \pi^2 , \; y =
0,  \; m_{\sigma} = 0.2  m_{\Phi} ,  \; \eta(0) = 1 $ and  $ {\dot
\eta}(0) = 0 $. 

Figure 8 shows a very rapid, non-exponential damping within few
oscillations of the expectation value and a saturation effect when the
amplitude of the oscillation is rather small (about 0.1 in this case), the
amplitude remains almost constant at the latest times tested. Figures 8 and
9 clearly show that the time scale for dissipation (from fig. 8) is that 
for which the particle production mechanism is more efficient
(fig. 9). Notice that the total number of particles produced rises on the
same time scale as that of damping in fig. 8 and eventually when the
expectation value oscillates with (almost) constant amplitude the average
number of particles produced remains constant. This behaviour is a
close analog to  the selfcoupled inflaton for unbroken symmetry (fig.1). 
The amplitude expansion predictions are in  qualitative
agreement with both results.

These figures clearly show that
damping is a consequence of particle production. At times larger than about 40
$m_{\Phi}^{-1}$ (for the initial values and couplings chosen) there is no
appreciable damping. The amplitude is rather small and particle production has
practically shut off. If we had used the {\it classical} evolution of the
expectation value in the mode equations, particle production would not shut off
(parametric resonant amplification), and thus we clearly see the dramatic
effects of the inclusion of the back reaction.

In ref.\cite{nos3} the broken symmetry case $ m^2_{\Phi} < 0 $ is then
studied. Figures 11-13  show $\eta(\tau)$ vs $\tau$,
${\cal{N}}_{\sigma}(\tau)$ vs 
$\tau$ and ${\cal{N}}_{q,\sigma}(\tau=200)$ vs $q$ respectively, for
$\lambda / 8\pi^2 = 
0.2;~~~ g / \lambda = 0.05;~~~ m_{\sigma}= 0.2\, |m_{\Phi}|; ~~
\eta(0)=0.6;~~~\dot{\eta}(0)=0$. Notice that the mass for the linearized
perturbations of the $\Phi$ field at the broken symmetry ground state is
$\sqrt{2}\,|m_{\Phi}| > 2 m_{\sigma}$. Therefore, for the values used in the
numerical analysis, the two-particle decay channel is open.
 For these values of the parameters, linear relaxation predicts
exponential decay with a time scale $\tau_{rel} \approx 300$ (in the units
used). Figure 11 shows very rapid non-exponential damping on time scales
about {\em six times shorter} than that predicted by linear relaxation. The
expectation value reaches very rapidly a small amplitude regime, once this
happens its amplitude relaxes very slowly. 
In the non-linear regime relaxation is clearly {\em
not} exponential but  extremely fast. The amplitude at long times
seems to relax to the expected value, shifted slightly from the
minimum of the tree level potential at $\eta = 1$. This is as expected
from the fact that there are quantum corrections. 
 Figure 12 shows that particle production occurs during the
time scale for which dissipation is most effective, giving direct proof that
dissipation is a consequence of particle production. Asymptotically, when the
amplitude of the expectation value is small, particle production shuts off. We
point out again that this is a consequence of the back-reaction in the
evolution equations. Without this back-reaction, as argued above, particle
production would continue without indefinitely. Figure 13 shows that the
distribution of produced particles is very far from thermal and concentrated at
low momentum modes $k \leq |m_{\Phi}|$. This distribution is qualitatively
similar to that in the unbroken symmetry case, and points out that the excited
state obtained asymptotically is far from thermal.

In ref.\cite{nos3} the case where the inflaton is only coupled to
fermions is studied ($g=0,\; y\neq 0$). The damping of the zero mode
is very inefficient in such case due to Pauli blocking. Namely, the
Pauli exclusion principle forbids the creation of more than $ 2 $
fermions per momentum state.  Pauli
blocking shuts off particle production and dissipation very early on. 

\section{Future perspectives}

The preheating and reheating theory in inflationary cosmology is
currently a very active area of 
research in  fast development, with the potential for dramatically
modifying the picture of the 
late stages of inflationary phase transitions. 

As remarked before, estimates and field theory calculations have
been done mostly  assuming Minkowski spacetime. Results in
de Sitter\cite{fut2}  and FRW backgrounds\cite{ult} are just beginning
to emerge. 
A further important step will be to consider the background
dynamics. Namely, the coupled gravitational and matter field
dynamics. The matter state equations obtained in Minkowski\cite{big}  and de
Sitter backgrounds\cite{fut}  give an indication, through the
Einstein-Friedmann equation, of the scale factor behaviour.

\clearpage

\hbox{\epsfxsize 14cm\epsffile{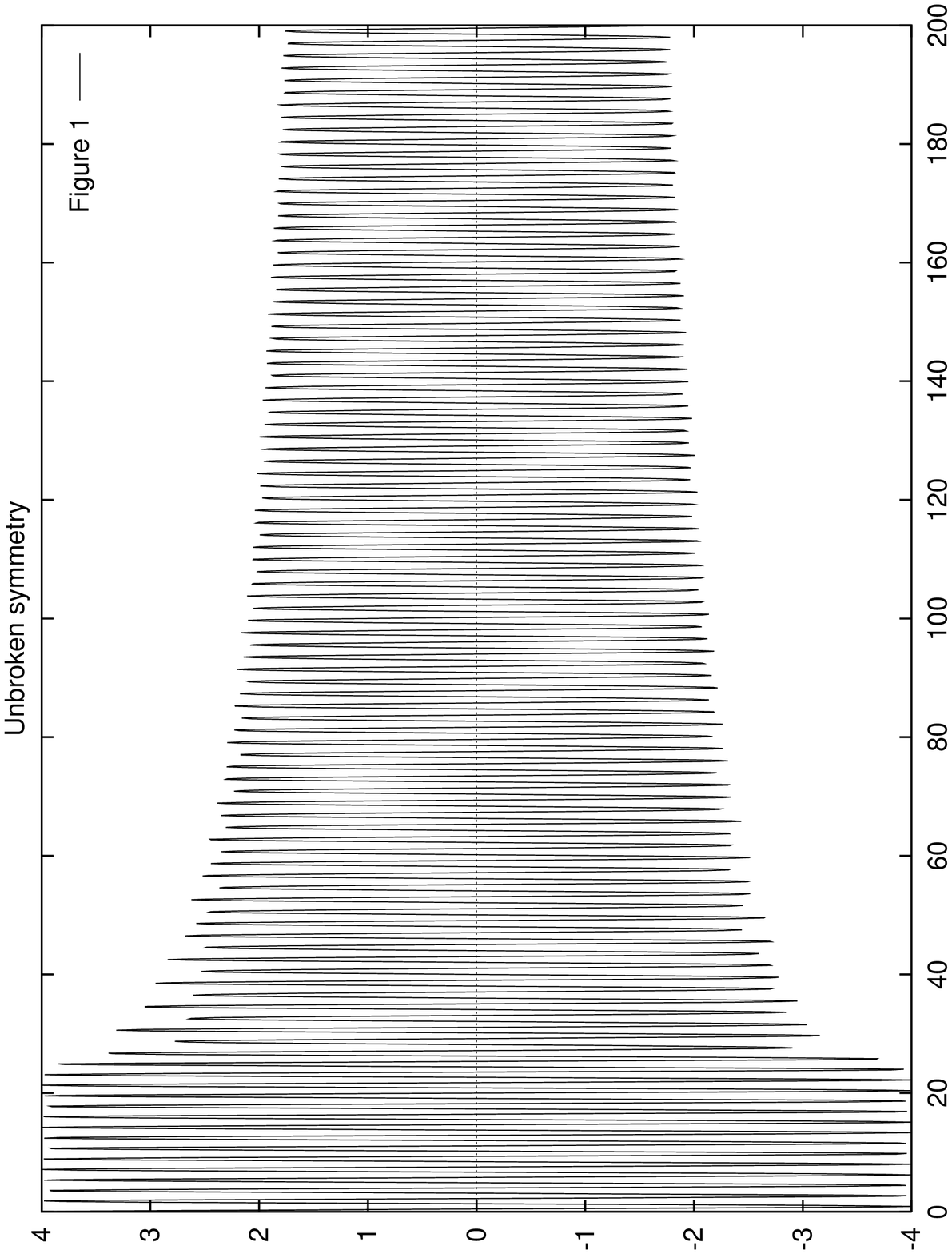}}

\figure{{\bf Figure 1:} 
$\eta(\tau)$ vs. $\tau$ for the unbroken symmetry case with
$\eta_0=4$, $g=10^{-12}$.\label{fig1}}

\clearpage

\hbox{\epsfxsize 14cm\epsffile{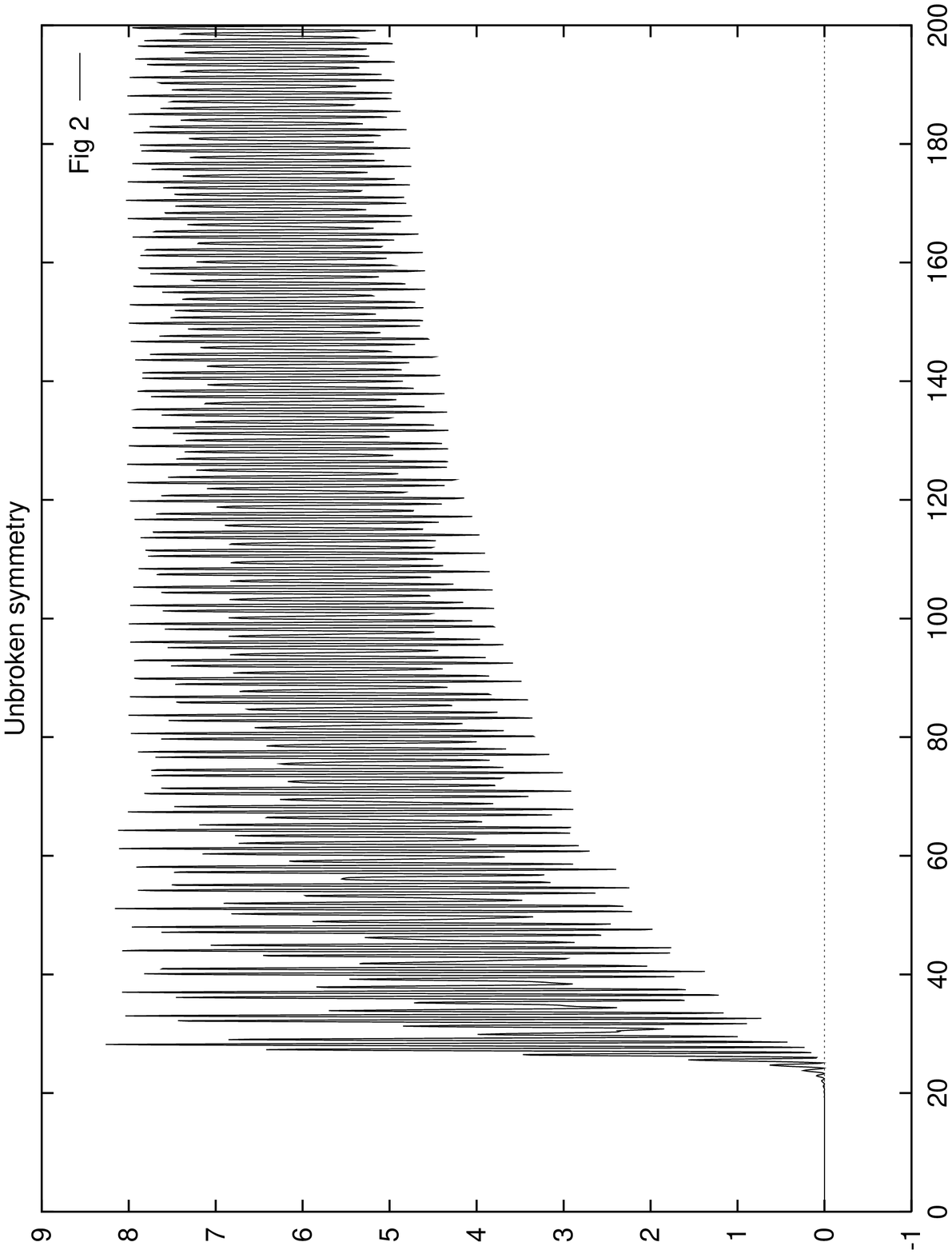}}

\figure{{\bf Figure 2:} 
$g\Sigma(\tau)$  vs. $\tau$ for the same values of the parameters as in
Fig. 1. \label{fig2}}

\clearpage

\hbox{\epsfxsize 14cm\epsffile{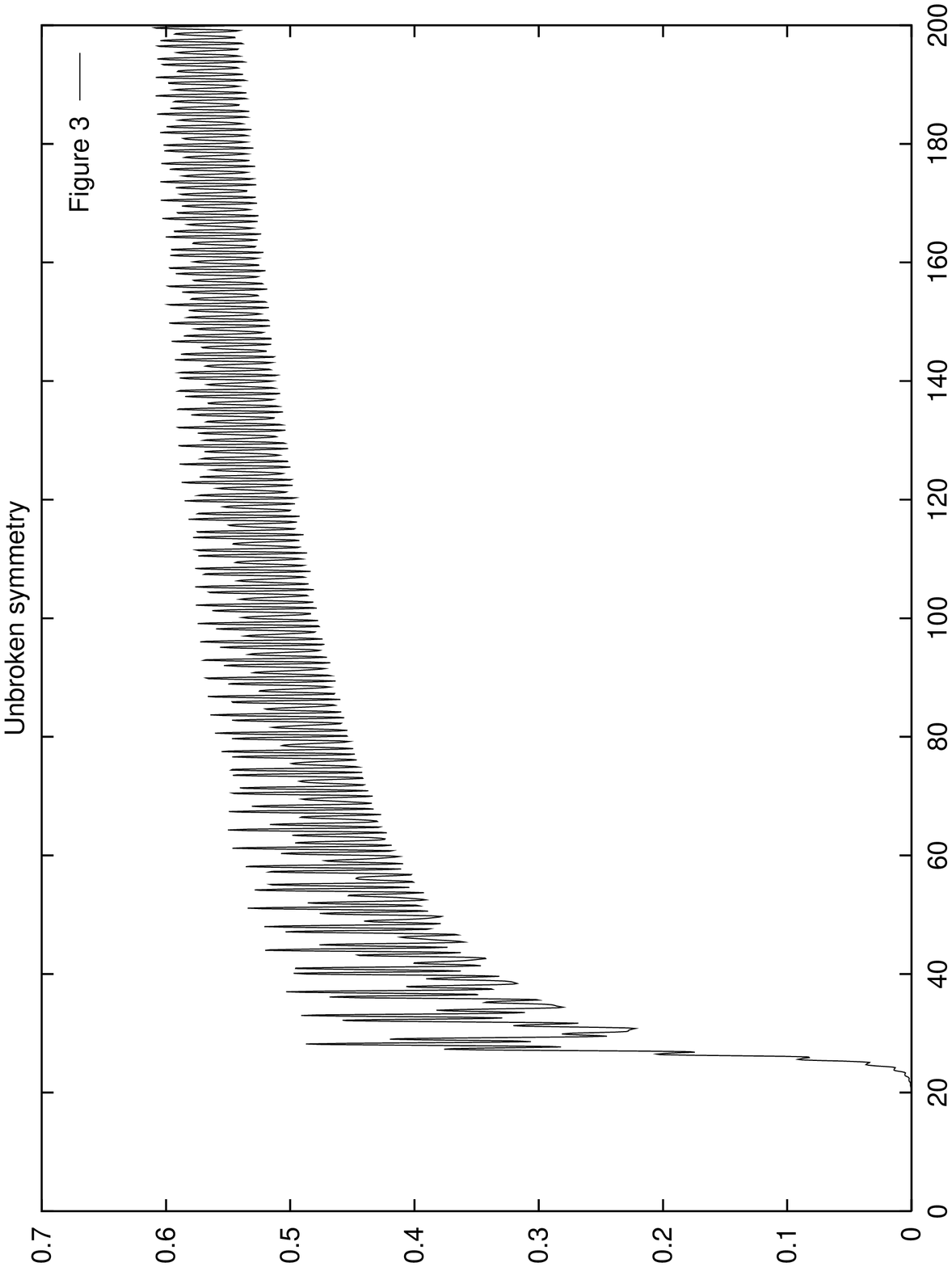}}

\figure{{\bf Figure 3:}
$g{\cal{N}}(\tau)$ for the same parameters as in
fig. 1.\label{fig3}}

\clearpage

\hbox{\epsfxsize 14cm\epsffile{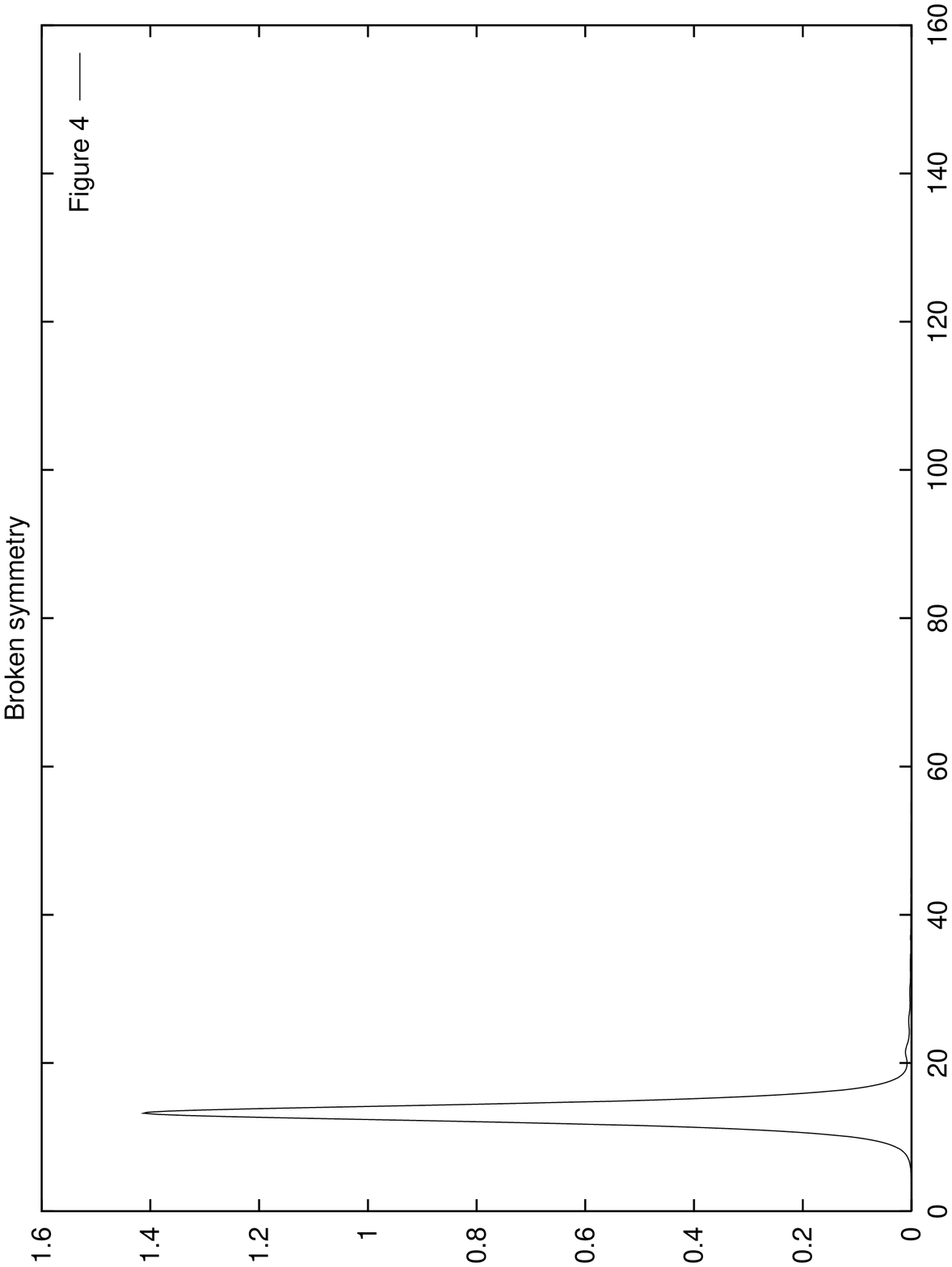}}

\figure{{\bf Figure 4:}
$\eta(\tau)$ vs. $\tau$ for the broken symmetry 
case with
$\eta_0=10^{-5}$, $g=10^{-12}$.\label{fig4}}

\clearpage

\hbox{\epsfxsize 14cm\epsffile{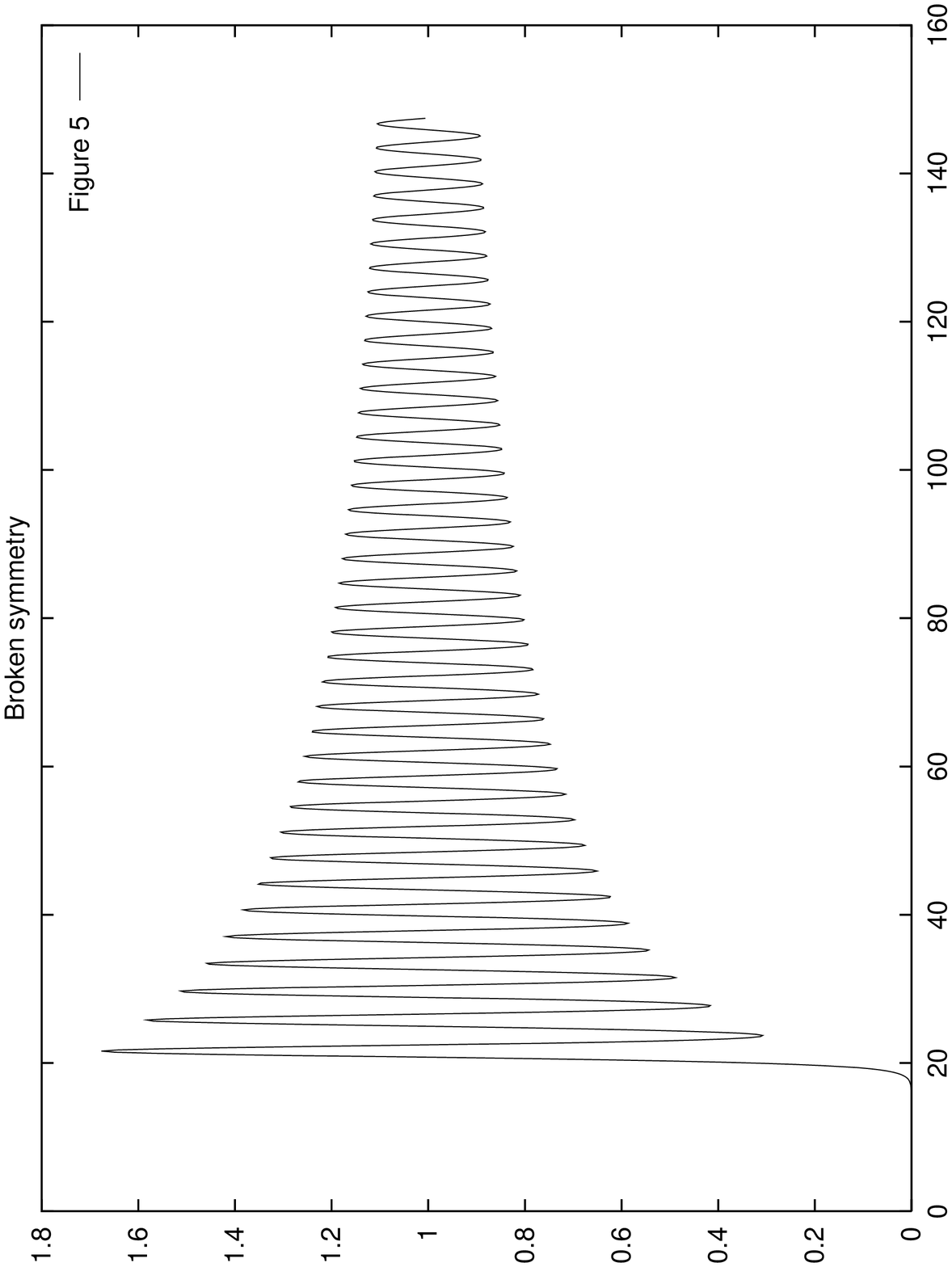}}

\figure{{\bf Figure 5:}
$g\Sigma(\tau)$ for the same values of the 
parameters as in fig. 4.\label{fig5}}

\clearpage

\hbox{\epsfxsize 14cm\epsffile{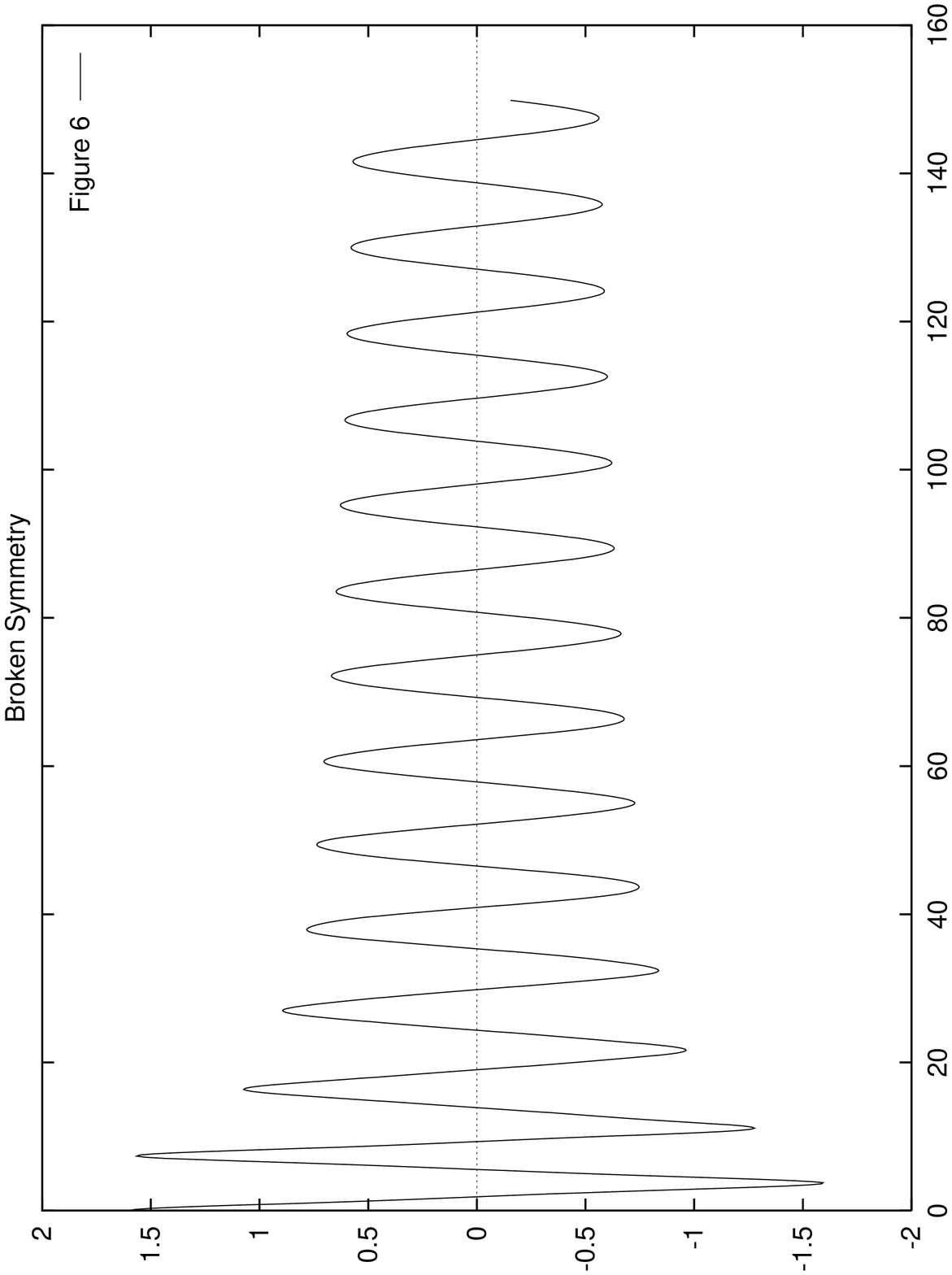}}

\figure{{\bf Figure 6:}
$\eta(\tau)$ vs. $\tau$ for the broken symmetry 
case with
$\eta_0=1.6 > \sqrt2 $, $g=10^{-3}$.\label{fig6}}

\clearpage

\hbox{\epsfxsize 14cm\epsffile{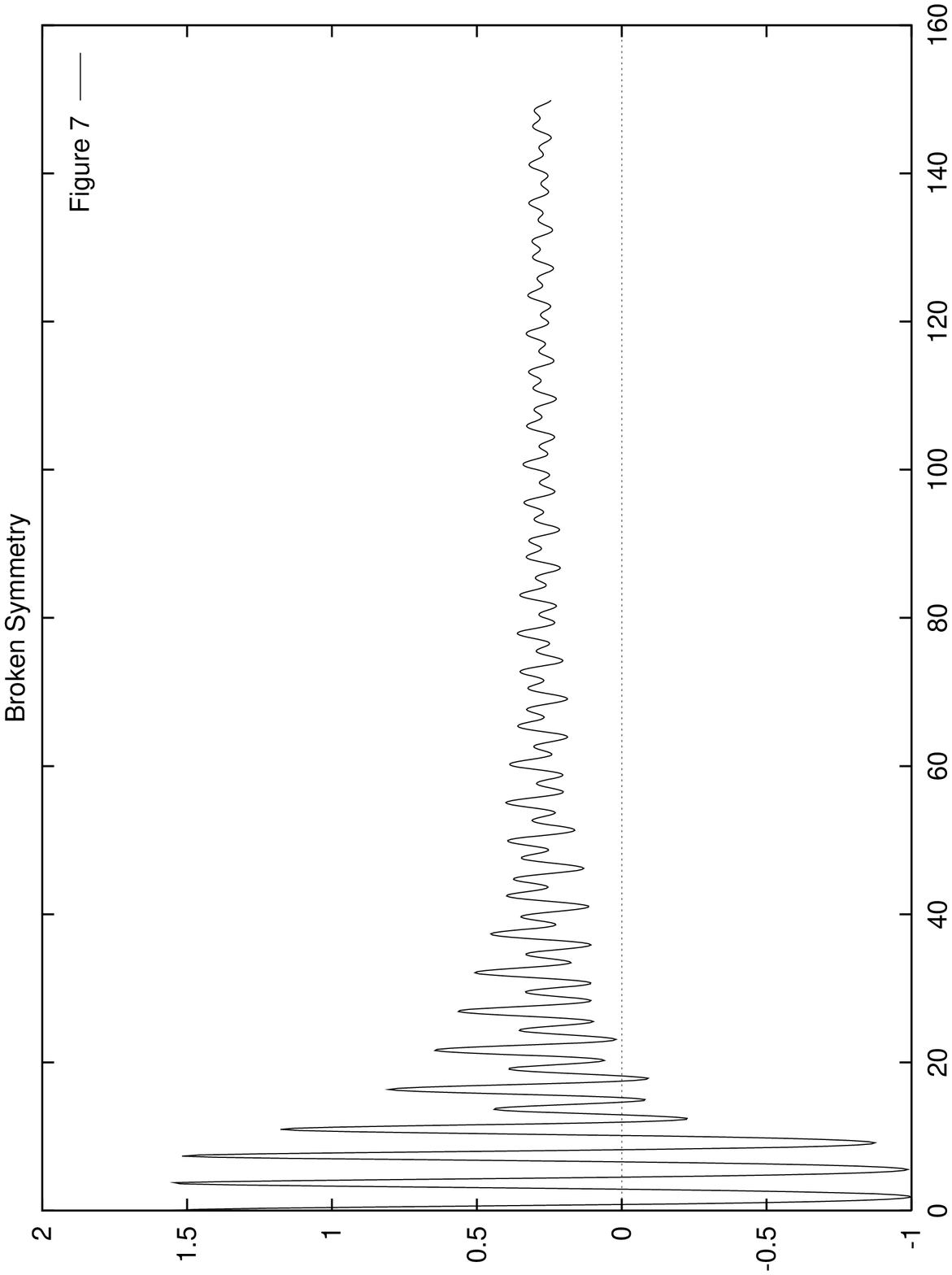}}

\figure{{\bf Figure 7:}
The effective mass squared  vs. $\tau$ for the same values of the 
parameters as in fig. 6.\label{fig7}}

\clearpage

\hbox{\epsfxsize 14cm\epsffile{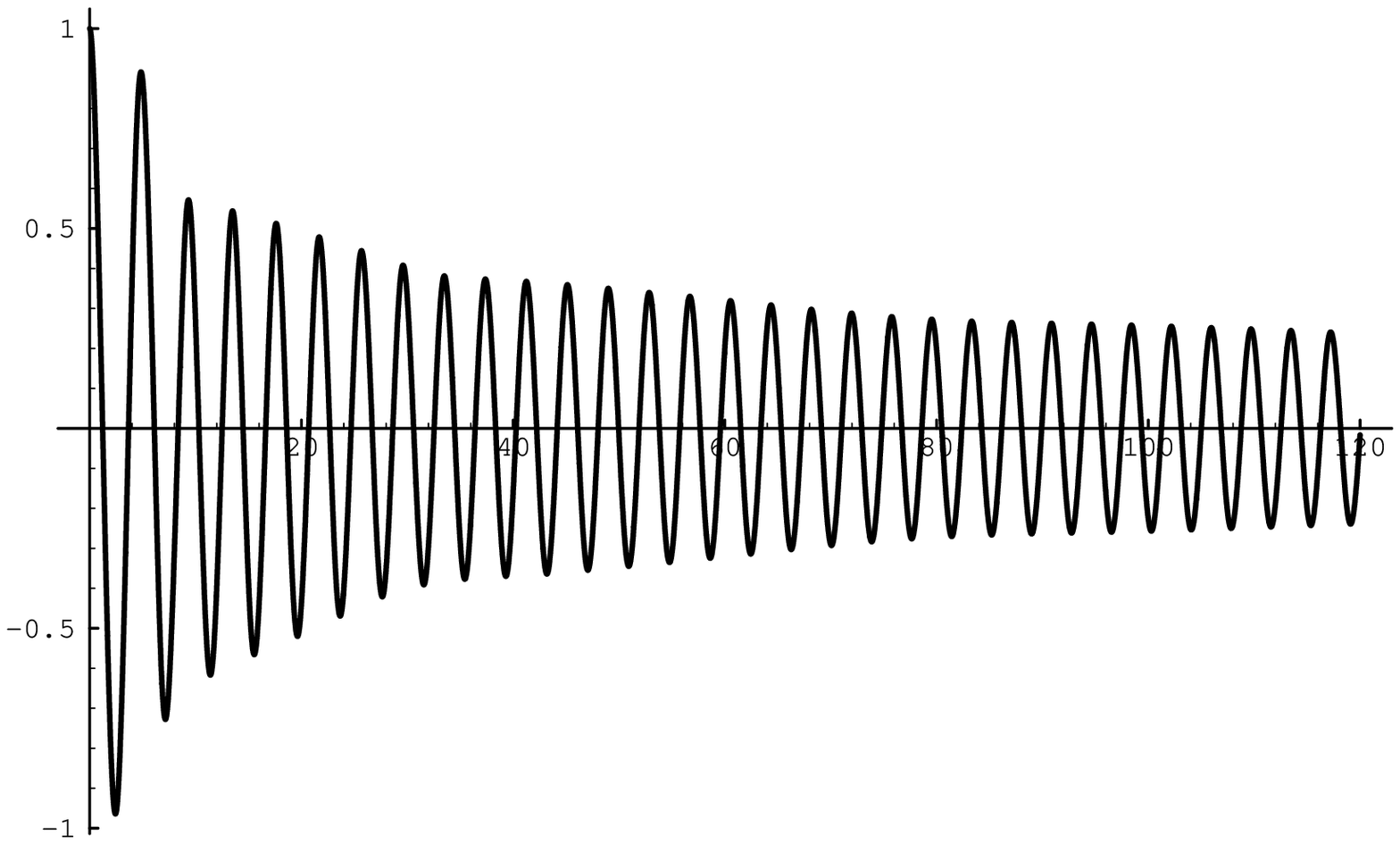}}

\figure{{\bf Figure 8:}
 The inflaton coupled to a lighter scalar field $\sigma$:
$\eta(\tau)$ vs $\tau$ for the values of the parameters
$y=0;~~ \lambda /8\pi^2=0.2;~~g=\lambda; ~
 m_{\sigma}=0.2\,m_{\phi};~~ \eta(0)=1.0;~~\dot{\eta}(0)=0$.  
\label{fig8}}

\clearpage

\hbox{\epsfxsize 14cm\epsffile{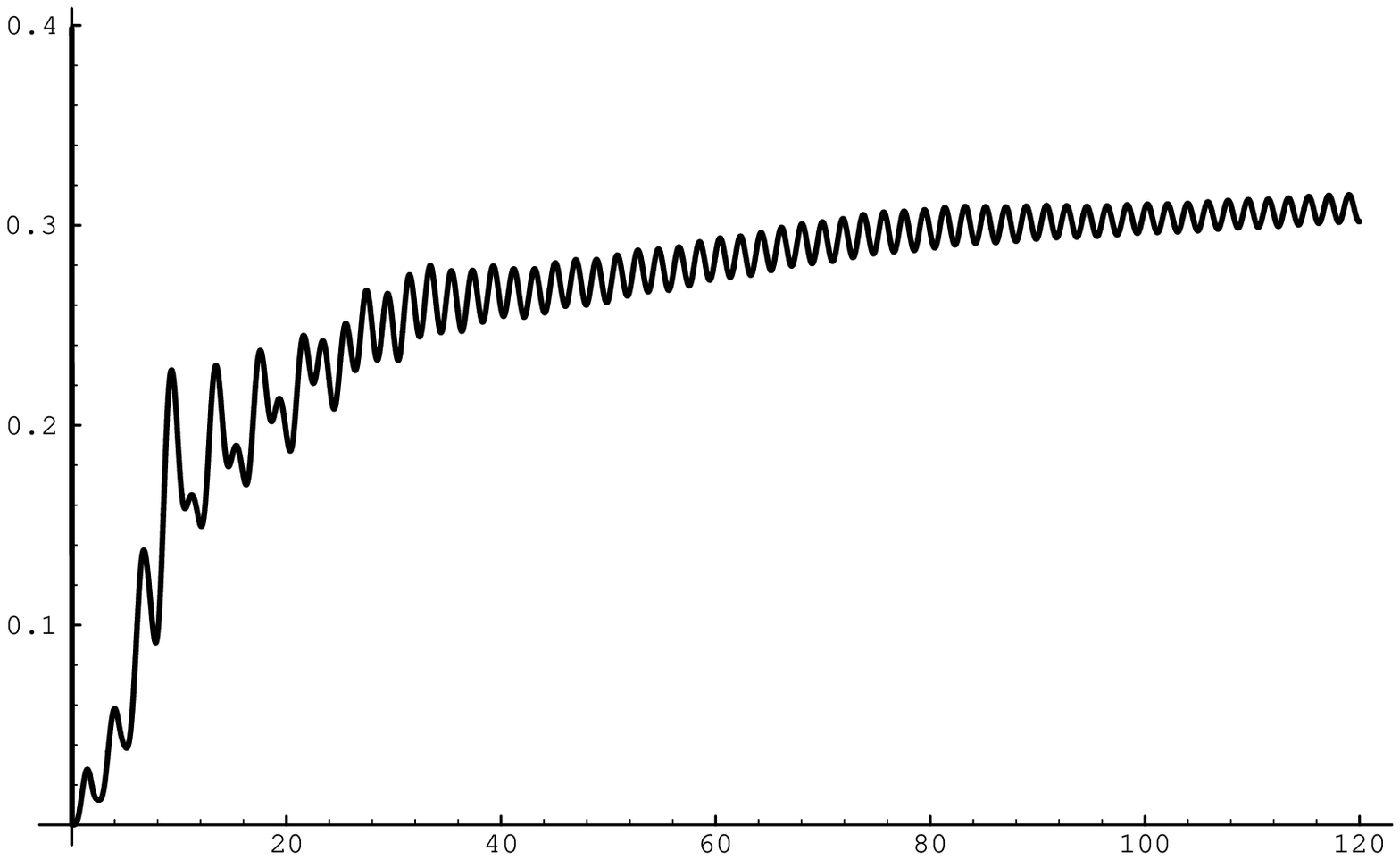}}

\figure{{\bf Figure 9:}
${\cal{N}}_{\sigma}(\tau)$ vs. $\tau$ for the same 
value of the parameters as figure 8. \label{fig9}}

\clearpage

\hbox{\epsfxsize 14cm\epsffile{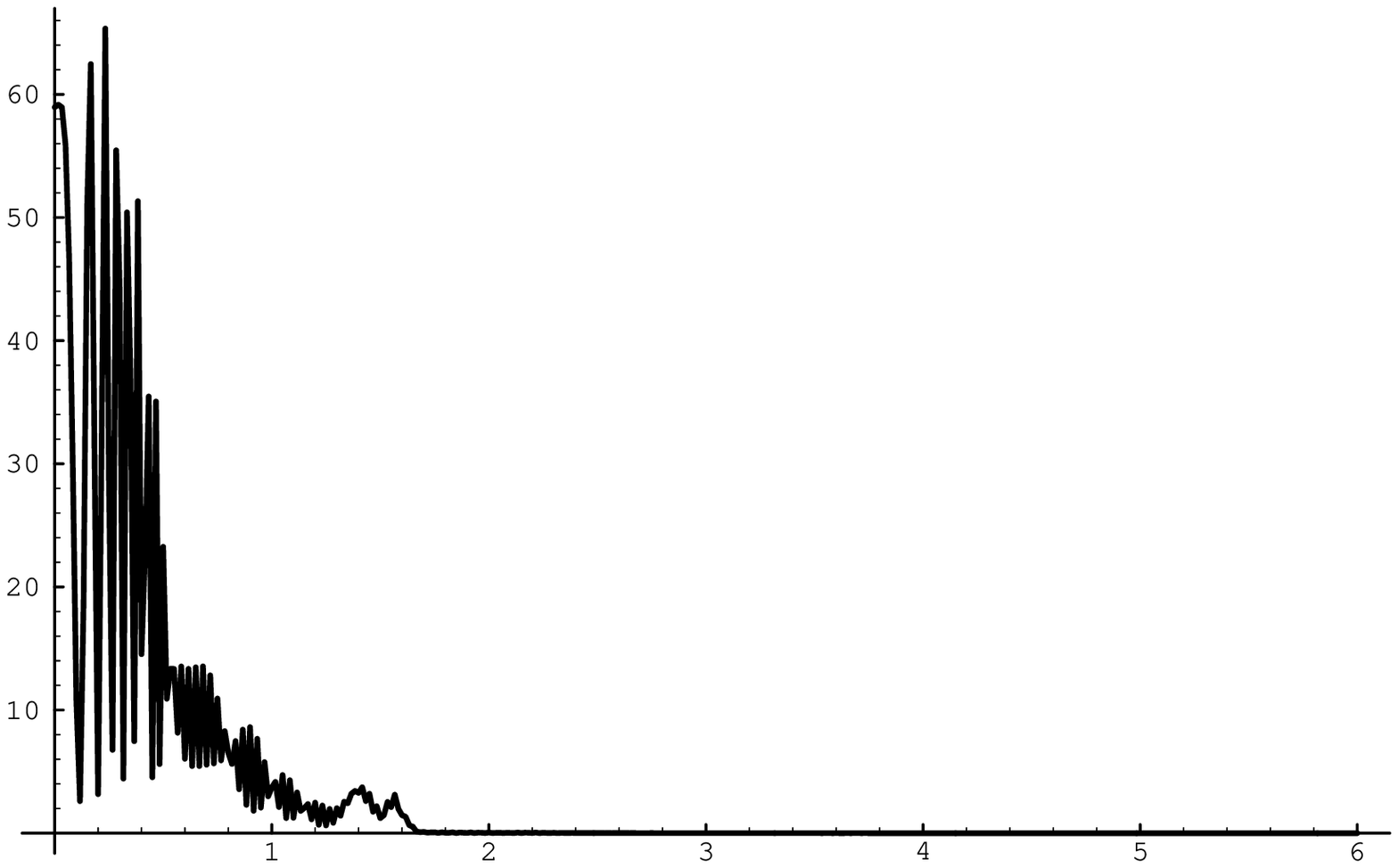}}

\figure{{\bf Figure 10:}
${\cal{N}}_{q,\sigma}(\tau=120)$ vs. $q$
for the same values as in fig. 8.  \label{fig10}}

\clearpage

\hbox{\epsfxsize 14cm\epsffile{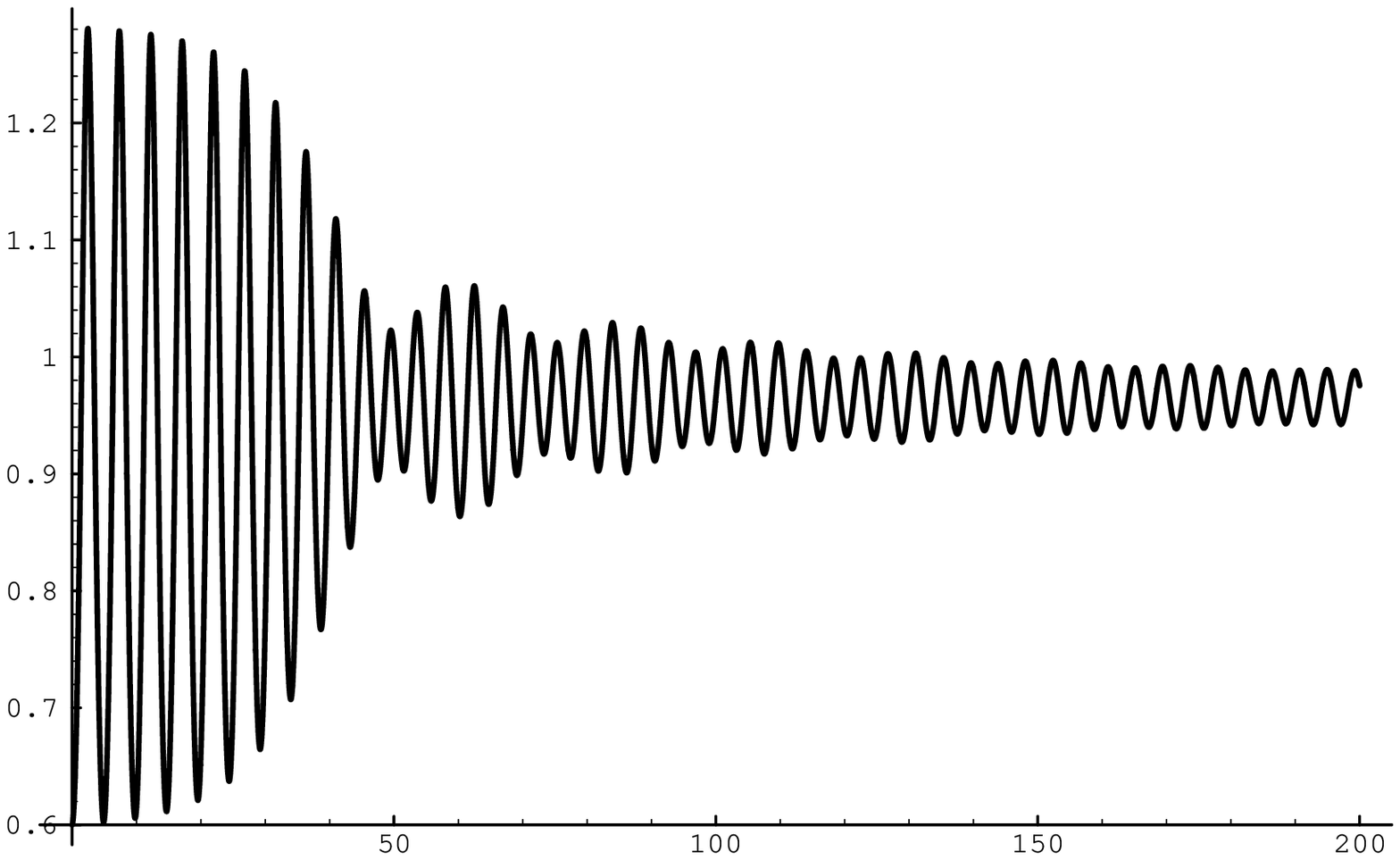}}

\figure{{\bf Figure 11:}
The inflaton in the  broken symmetry case
coupled to a lighter scalar $\sigma$.
$\eta(\tau)$ vs $\tau$ for the values of the parameters
$y=0;~~ \lambda /8\pi^2=0.2;~~g=\lambda; ~
 m_{\sigma}=0.2\,|m_{\phi}|;~~ \eta(0)=0.6;~~\dot{\eta}(0)=0$.
 \label{fig11}}

\clearpage

\hbox{\epsfxsize 14cm\epsffile{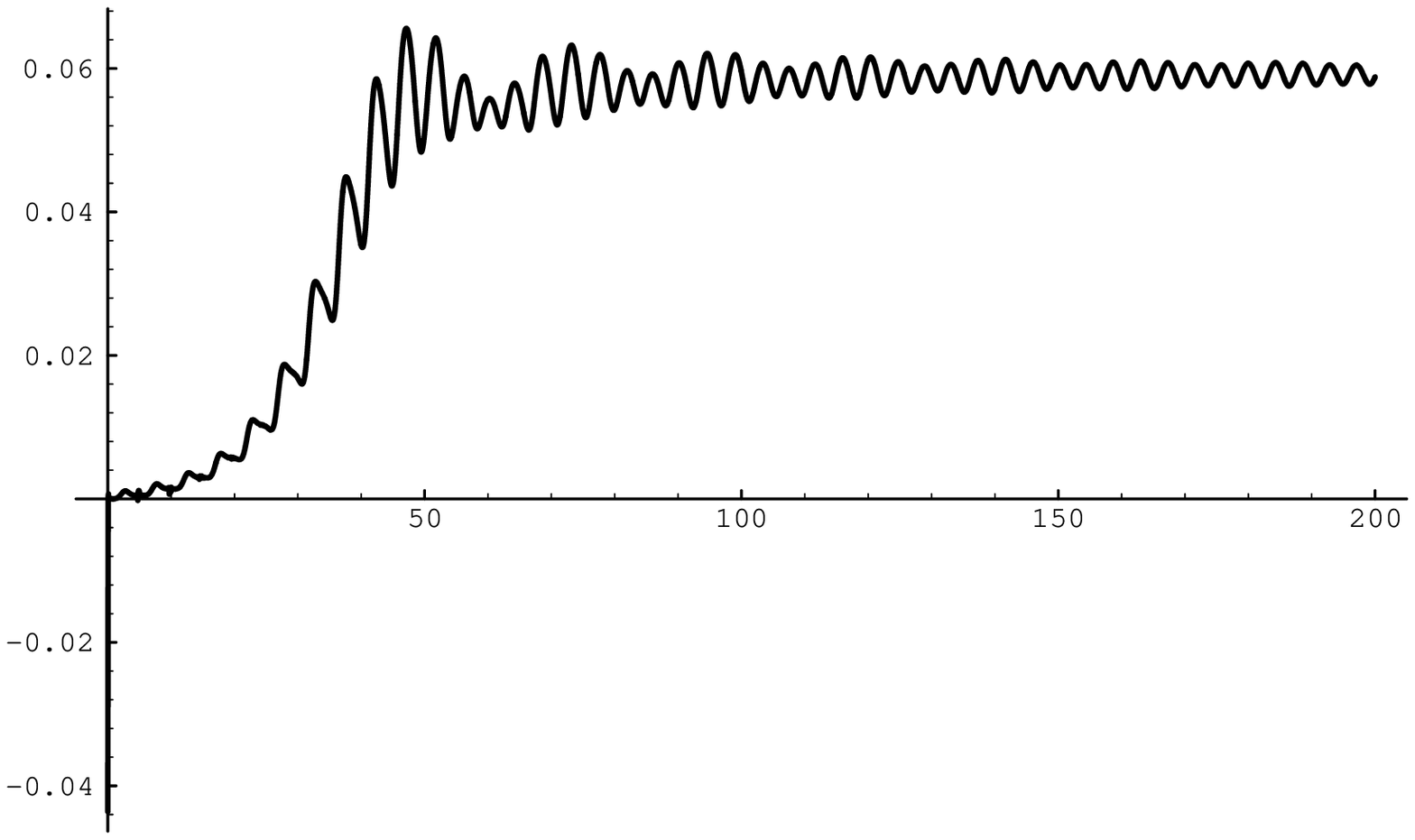}}

\figure{{\bf Figure 12:}
${\cal{N}}_{\sigma}(\tau)$ vs. $\tau$ for the same 
value of the parameters as fig. 11.  \label{fig12}}

\clearpage

\hbox{\epsfxsize 14cm\epsffile{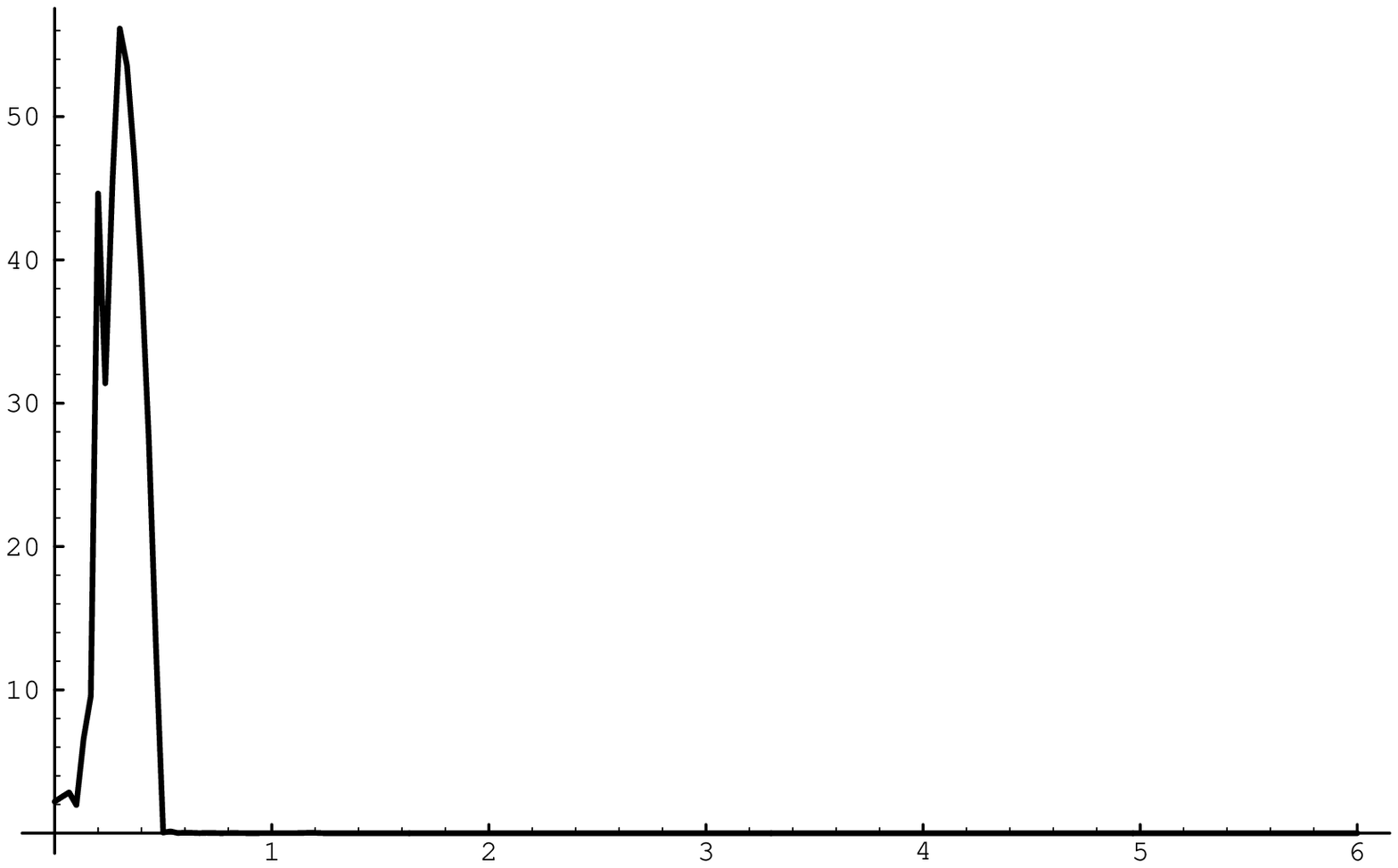}}

\figure{{\bf Figure 13:}
${\cal{N}}_{q,\sigma}(\tau=200)$ vs. $q$
for the same values as in fig. 11.  \label{fig13}}

\end{document}